\documentclass[12pt]{iopart}
\usepackage{graphicx}

\newcommand{\dgg}{^{\dagger}}
\newcommand{\nn}{\nonumber}
\def\parf{\stackrel{\leftrightarrow}{\partial_i}}
\def\tr{\mbox{tr}\{}

\begin{document}

\title[First correction to
JIMWLK evolution from the classical EOMs]{First correction to
JIMWLK evolution from the classical equations of motion}

\author{J L Albacete$^1$, \underline{N Armesto}$^2$ and  J G Milhano$^3$}

\address{$^1$ Department of Physics, The Ohio State University,
        191 W. Woodruff Avenue, 43210 Columbus, OH, USA\\
$^2$ Departamento de F\'{\i}sica de Part\'{\i}culas and
IGFAE,
Universidade de Santiago de Compostela,
 15782 Santiago de Compostela, Spain\\
$^3$ CENTRA, Instituto Superior T\'ecnico (IST),
Av. Rovisco Pais, P-1049-001 Lisboa, Portugal}
\ead{$^1$ albacete@mps.ohio-state.edu, $^2$ nestor@fpaxp1.usc.es, $^3$
gui@fisica.ist.utl.pt}

\begin{abstract}
We calculate some ${\cal O}(\alpha_s^2)$
corrections to the JIMWLK kernel in the
framework of the light-cone wave function approach
to the high energy limit of
QCD. The contributions that we consider originate from higher order
corrections in the strong coupling and in the density of the projectile to
the solution of the classical Yang-Mills equations of motion
that determine the Weizs\"acker-Williams
fields of the projectile. We study the structure of these corrections in the
dipole limit, showing that they are subleading in the limit of large
number of colours $N$, and that
they cannot be fully recast in the form of dipole degrees of freedom.
\end{abstract}


\section{Introduction: the wave function formalism and JIMWLK}
\label{intro}

The study of high gluon density QCD, pioneered in the MV model
\cite{McLerran:1993ni}, has become a fashionable subject both due to its
theoretical relevance and to eventual applications to small-$x$
DIS or high-energy nuclear collisions (see the review \cite{Iancu:2003xm}). The
main outcome of these studies was the JIMWLK equation which controls the
evolution of the colour distribution of a hadron with increasing
energy. In the last 3 years, the limitations of such evolution equation have
become apparent in the form of the discussion of fluctuations and pomeron
loops, see the reviews \cite{Kovner:2005pe}.
In all available schemes, a separation of scales is considered
between fast partons which act as sources for the classical dynamics of the
soft, slow
glue.
Considering gluon radiation and the arbitrariness
of the fast-slow separation led to a renormalization group equation, the
JIMWLK equation. A powerful framework to derive this equation, in which its
limitations become apparent, is the wave function approach (WFA)
\cite{Kovner:2005jc} which we briefly introduce in this Section,
see full details in \cite{Kovner:2005jc,Albacete:2006uv}.

In the WFA the evolution is introduced in the wave function of the projectile
scattering off a hadronic target. This wave function is the superposition of
fast gluons with $k^+>\Lambda$. The scattering matrix with the target is the
superposition of independent
scattering matrices $S^{ab}$.
Then, an average over target configurations is taken,
\begin{equation}
\Sigma_Y[S]=\langle \Psi^i(Y)|\Psi^f(Y)\rangle, \qquad
\langle \Sigma \rangle
= \int dS \,\Sigma[S]\,W_Y[S],
\end{equation}
with $W_Y[S]$ the probability
density  for the target to
have a certain configuration of the fields.
After a boost $Y \to Y+\delta Y$, the wave functions become
\begin{eqnarray}
&&|\Psi^f(Y+\delta Y)\rangle = \left\{1-\frac{1}{2}\delta Y\int
d^2z\,b_i^a(z,S\rho)b_i^a(z,S\rho)\right.
\label{eq2} \\
& &+  \left. i\int d^2z\,b_i^a(z,S\rho)\int_{(1-\delta
Y)\Lambda}^{\Lambda}\frac{dk^+}{\sqrt{\pi}|k^+|^{1/2}}S^{ab}(z)a_i^{\dgg
b}(k^+,z)\right\}|\Psi^f(Y)\rangle,\nonumber
\end{eqnarray}
with $S^{ab} \to \delta^{ab}$ for $|\Psi^i(Y+\delta Y)\rangle$, $\rho$ the
projectile density operator and $b_i^a$ the Weizs\"acker-Williams
fields of the projectile (WW)
determined from the
classical Yang-Mills equations of motion (EOMs),
in which the hard gluons enter as an external
source.

\begin{figure}[htb]
\begin{center}
\vskip -0.cm
\includegraphics[width=13cm,height=4.2cm]{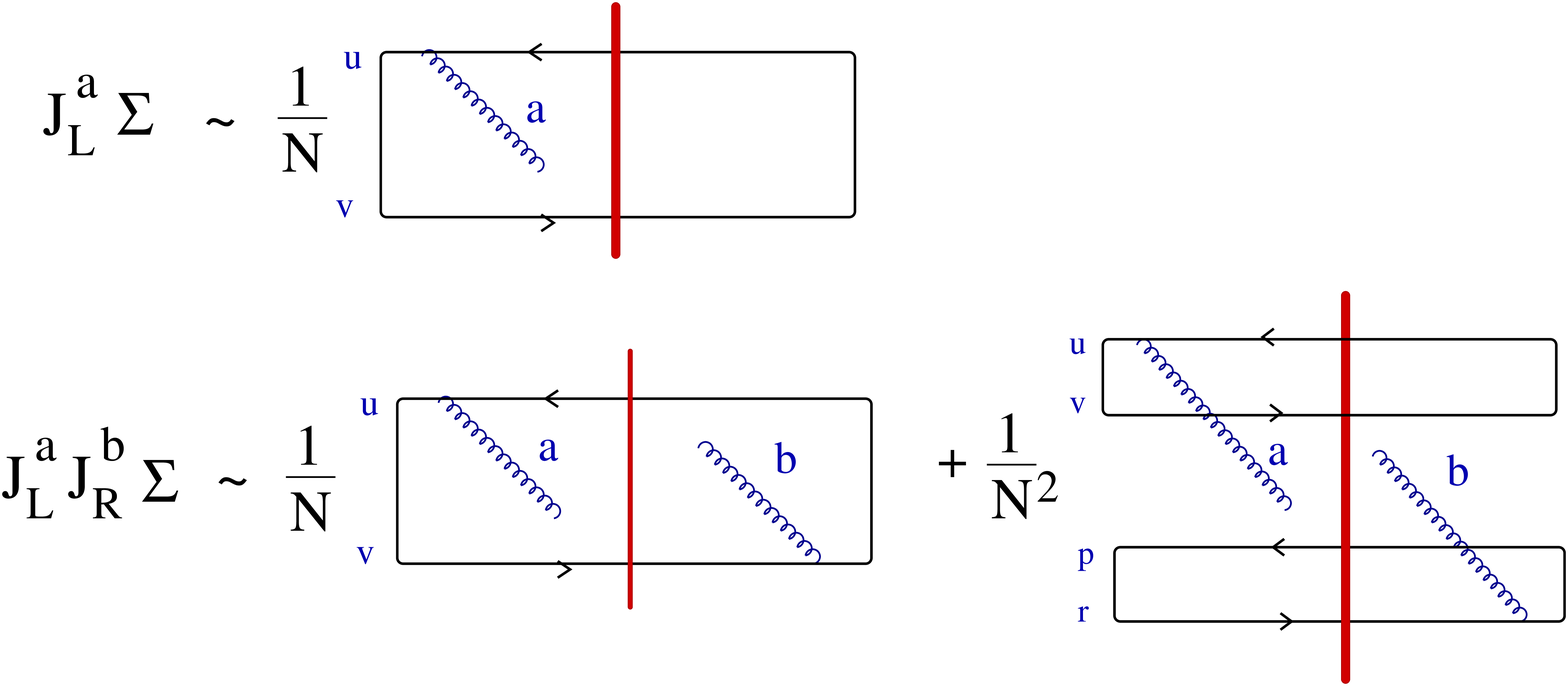}
\vskip -0.cm
\caption{Diagrammatic representation of $J_L^a\Sigma$ (top) and
$J_L^aJ_R^b\Sigma$ (bottom). The target is represented by the
vertical thick line.}
\label{fig1}
\end{center}
\end{figure}

The
resulting evolution equation is $\delta \Sigma[S]/\delta Y=\chi\, \Sigma[S]$,
with the kernel given by
\begin{equation}
\chi=-\frac{1}{2\pi}\int_z\,\left[b_i^a(J_L)b_i^a(J_L)+
b_i^a(J_R)b_i^a(J_R)-2S^{ba}(z)b_i^a(J_L)b_i^b(J_R)
\right],
\label{kerb}
\end{equation}
where $b_i^a$ depends on $z$ and $J_{L(R)}$ are left (right) colour rotation
operators, whose action in the dipole model limit, with the projectile
composed of colour dipoles \cite{Kovner:2005jc}, is illustrated
in \Fref{fig1}.
With the EOMs solved at lowest order in $g\rho$
in the $A^+=0$ gauge
\cite{McLerran:1993ni},
\begin{equation}
b^a_i(z)=g\left(b^a_{i1}(z)+g^2\,b^a_{i2}(z)+\mathcal{O}(g^4)\right), \qquad
b^a_{i1}(z)=
\int \!{d^2x\over 2 \pi}\,\left(\partial_iX\right)\rho^a(z),
\label{o1}
\end{equation}
with
$\partial_iX\equiv \partial^{z}_i \ln\left(|z-x|\lambda\right)$, the JIMWLK
evolution equation is obtained \cite{Kovner:2005jc}.

\vskip -0.5cm
\section{First correction from the classical equations of motion}
\label{corr}

Two limitations of the derivation leading to JIMWLK are apparent: First,
\Eref{eq2} is an approximation only
valid for a low-density projectile. Second,
the solution of the EOMs is done at the lowest order in the density of the
projectile, \Eref{o1}. In this Section we explore the corrections to the
latter by going to order $g^3\rho^2$
\cite{Albacete:2006uv,Kovchegov:1996ty},
\begin{equation}
b^a_{i2}(z)=-\frac{1}{4(2\pi)^2}f^{abc}\!\int d^2x\,d^2y\,(X\!\parf
\!Y)\rho^b(x)\rho^c(y),
\end{equation}
$X\!\parf\! Y\equiv X\left(\partial^{z}_iY\right)
-\left(\partial^{z}_iX \right)Y$.
With this expression the expansion of the kernel becomes
\begin{equation}
\chi=\alpha_s\left(\chi^{1)}+\alpha_s\,\chi^{2)}+\mathcal{O}(\alpha_s^2)
\right),
\label{expk}
\end{equation}
whose leading term is the JIMWLK kernel
\begin{equation}
\alpha_s\chi^{1)}= -\frac{\alpha_s}{2\pi^2}\int_{xyz}
\!\!(\partial_iX)(\partial_iY)\left[J_L^aJ_L^a+
J_R^aJ_R^a-2S^{ba}J_L^aJ_R^b\right],
\label{kjimwlk}
\end{equation}
with the $J$'s and $S^{ab}$ depending on $z$,
and the first correction reads
\begin{eqnarray}
&&\chi^{2)}=-\frac{1}{(2\pi)^2}\int_{xywz}f^{abc}\,(\partial_i Y)
(X\!\parf\! W)\,
\left[J_L^a(x)J_L^b(y)J_L^c(w)+{\rm RRR}\right]\nonumber  \\
& &-2f^{acd}(\partial_i X)(Y\!\parf\! W)S^{ba}(z)\left[J_L^b(x)J_R^c(y)
J_R^d(w)+{\rm LLR}\right].
\label{o3}
\end{eqnarray}

\begin{figure}[htb]
\begin{center}
\vskip -0.cm
\includegraphics[width=13cm,height=7.8cm]{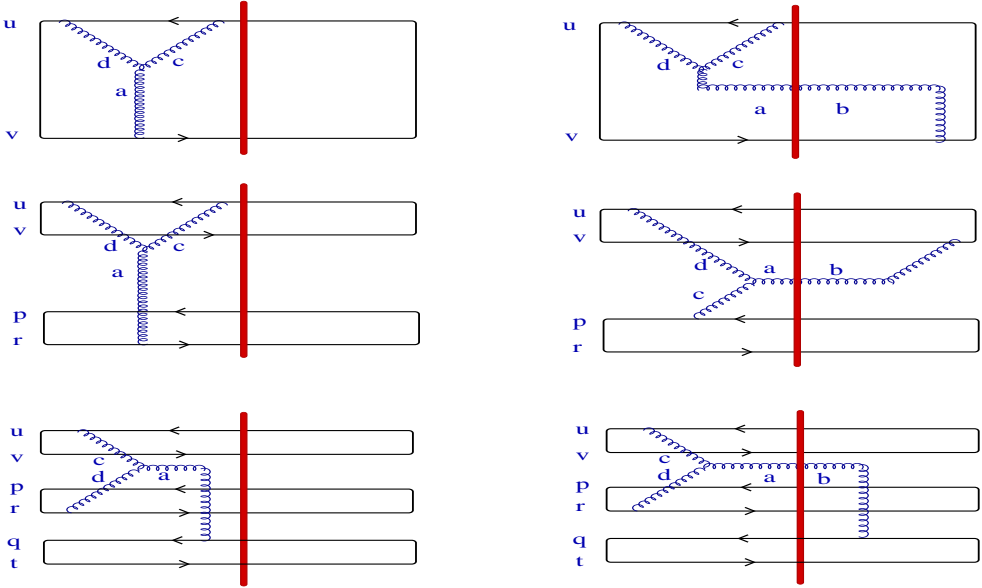}
\vskip -0.cm
\caption{Diagrams corresponding to LLL (plots on the left) and LLR
(plots on the right), see the text. The
diagrams include one ($\propto 1/N$, top), two ($\propto 1/N^2$,
middle) and three  ($\propto 1/N^3$, bottom) active
dipoles.}
\label{fig2}
\end{center}
\end{figure}

To proceed further we go to the dipole model limit, in which the action of the
different terms can be explicitly computed and is illustrated in \Fref{fig2}.
With $s(x,y)$ the $S$-matrix for a dipole, the kernel
(\ref{o3})
can be separated \cite{Albacete:2006uv}
into three pieces containing one, two or
three active dipoles with the corresponding $1/N$ suppression factors,
\begin{equation}
\chi^{2)}\Sigma[s]=\left(\chi^{2)}_{1/N}+\chi^{2)}_{1/N^2}+\chi^{2)}_{1/N^3}
\right)\Sigma[s].
\end{equation}
$\chi^{2)}_{1/N}$ is found to vanish. Thus the solution to
next-to-leading
order in the EOMs does not yield leading $1/N$ corrections and,
therefore,
it does not provide a correction to the BK equation
\cite{Balitsky:1995ub}.
$\chi^{2)}_{1/N^2}$ and $\chi^{2)}_{1/N^3}$
cannot be fully recast in terms of dipoles. Noting that any wave function or
weight functional of a gluonic/dipole configuration has to be completely
symmetric under the exchange of any number of gluons/dipoles, we get
\begin{eqnarray}
& &\chi^{2)}_{1/N^2}\Sigma[s]=-\frac{i}{(2\pi)^2} \,\frac{1}{N^2}
\int d^2z\,\Big{\{} [\partial_i(R\!-\!P) (U\!\parf\! V)
\label{n2}
\\
& &-
\partial(V\!-\!U) (P\!\parf\!R)] [
N\,\tr S_u^\dagger S_v S_z^\dagger S_rS_p^\dagger S_z \}]
+\partial_i(V\!-\!U)
\nn
\\
& &
\times
(R\!-\!P)\parf (V\!-\!U)
\left[ \tr S_z^\dagger S_v S_p^\dagger S_r \} \tr  S_u^\dagger
S_z\}-\tr S_u^\dagger S_z S_p^\dagger S_r\} \tr  S_z^\dagger S_v\}\right.
\nn \\
& &\left.
 +\tr S_p^\dagger S_zS_u^\dagger S_r\}\tr  S_z^\dagger S_v\}
-\tr  S_p^\dagger S_vS_z^\dagger S_r\}\tr  S_u^\dagger
S_z\}\right]\Big{\}}\,\frac{\delta^2\Sigma[s]}{\delta s(p,r)\delta s(u,v)}\,\,,
\nn
\end{eqnarray}
\begin{eqnarray}
& &\chi^{2)}_{1/N^3}\Sigma[s]=-\frac{i}{2}\,\frac{1}{(2\pi)^2} \,\frac{1}{N^3}
\int d^2z \left[ \{\partial_i(R\!-\!P)\}
\{(T\!-\!Q)\parf (V\!-\!U)\}\right]
\nn \\
& &\times \left[\tr S^\dagger_uS_vS^\dagger_qS_tS^\dagger_pS_r\}
+\tr S^\dagger_uS_tS^\dagger_qS_rS^\dagger_pS_v\} +
\tr S^\dagger_zS_rS^\dagger_pS_zS^\dagger_uS_vS^\dagger_qS_t\}
\right.
\nn \\
& &\left.
+\tr S^\dagger_pS_rS^\dagger_zS_vS^\dagger_uS_t S^\dagger_qS_z\}\right]
\frac{\delta^3\Sigma[s]}{\delta s(q,t)\delta s(p,r)\delta s(u,v)}\,\,.
\label{n3}
\end{eqnarray}

\section{Conclusions}
\label{disc}

In this contribution we present the ${\cal O}(\alpha_s^2)$ corrections to the
JIMWLK evolution equation coming from the ${\cal O}(g^3\rho^2)$ solution to
the classical EOMs. The leading $1/N$ piece in the dipole model limit is found
to vanish, thus yielding no correction to the BK equation. Subleading
corrections do not show a closed dipole form. While the
corrections we compute are certainly not the complete set of ${\cal
O}(\alpha_s^2)$ corrections to JIMWLK, they are part of the full solution and
may turn to be important to fulfill general requirements
of the complete theory of high gluon density QCD like dense-dilute duality
\cite{Kovner:2005en}.

\ack
JLA, NA and JGM acknowledge 
financial support by the U.S. Department of Energy
under Grant No. DE-FG02-05ER41377, by Ministerio de Educaci\'on
y Ciencia of Spain under a contract Ram\'on y Cajal and project FPA2005-01963
and by Xunta de Galicia (Conseller\'{\i}a de Educaci\'on), and by the Funda\c
c\~ao para
a Ci\^encia e a Tecnologia of Portugal under contract
SFRH/BPD/12112/2003, respectively. NA thanks
the organizers for such a
nice conference.

\end{document}